\pdfoutput=1
\documentclass[a4paper,american,floatfix,pdftex,superscriptaddress,twocolumn,%
aps,
reprint%,final% add final to see real layout, no todonotes anymore, ...
]{revtex4-2}
\PassOptionsToPackage{unicode}{hyperref} % please leave linkframes in the file
\usepackage{amsfonts,amsmath,amssymb}
\usepackage{graphicx}
\usepackage[T1]{fontenc}
\usepackage[utf8]{inputenc}
\usepackage{hypernat}
\usepackage[ams]{CMImacros}
\usepackage{xr-hyper} % add cross-referencing to supplementary information

\newcommand*\ToT[1]{\ensuremath{\text{ToT}_\text{#1}}\xspace}
\newcommand*\ToTZ[1]{\ensuremath{\text{ToT}_\text{#1}^{(0)}}\xspace}
\newcommand*\IKrum{\ensuremath{I_\text{Krum}}\xspace}

\graphicspath{{./fig/}} % define figure directory
\usepackage{style}
\myexternaldocument[esi-]{build}{\jobname_SI}

\newcommand{\cfeldesy}{\affiliation{Center for Free-Electron Laser Science CFEL, Deutsches
      Elektronen-Synchrotron DESY, Notkestr.~85, 22607 Hamburg, Germany}}%
\newcommand{\uhhcui}{\affiliation{Center for Ultrafast Imaging, Universität Hamburg, Luruper
      Chaussee~149, 22761 Hamburg, Germany}}%
\newcommand{\uhhphys}{\affiliation{Department of Physics, Universität Hamburg, Luruper Chaussee~149,
      22761 Hamburg, Germany}}%
\newcommand{\desy}{\affiliation{Deutsches Elektronen-Synchrotron DESY, Notkestr.~85, 22607 Hamburg,
      Germany}}%
\newcommand{\cern}{\affiliation{CERN, Experimental Physics Department, Meyrin, 1211, Switzerland}}%
\newcommand{\stemail}{\email[Email:~]{sebastian.trippel@cfel.de}}%
\newcommand{\cmiweb}{\homepage[\\Website:~]{https://www.controlled-molecule-imaging.org}}%

\begin{document}
\title{Timepix3: single-pixel multi-hit energy-measurement behavior}%
\author{Hubertus Bromberger}\cfeldesy%
\author{David Pennicard}\desy %
\author{Rafael Ballabriga}\cern%
\author{Sebastian Trippel}\stemail\cmiweb\cfeldesy\uhhcui%
\author{Jochen Küpper}\cfeldesy\uhhcui\uhhphys%

\begin{abstract}
   \noindent%
   The event-driven hybrid-pixel detector readout chip, Timepix3, has the ability to simultaneously
   measure the time of an event on the nanosecond timescale and the energy deposited in the sensor.
   However, the behaviour of the system when two events are recorded in quick succession of each
   other on the same pixel was not studied in detail previously. We present experimental measurements, circuit
   simulations, and an empirical model for the impact of a preceding event on this energy
   measurements, which can result in a loss as high as 70~\%. Accounting for this effect enables
   more precise compensation, particularly for phenomena like timewalk. This results in significant
   improvements in time resolution -- in the best case, multiple tens of nanoseconds -- when two
   events happen in
   rapid succession.
\end{abstract}
\date{\today}%
\maketitle%

\section{Introduction}
In the rapidly evolving field of scientific and industrial imaging, there is an ever-increasing need
for detectors capable of sensing individual events with high temporal and spatial resolution, while
being able to cope with high event rates~\citep{Garcia-Sciveres:RPP81:066101, Wermes:nima924:44,
   Ponchut:radmea140:106459, Heijne:radmea140:106436}. Such detectors have a broad range of
applications, including nuclear medicine, high-speed industrial inspection, particle and atomic,
molecular, and optical (AMO)-physics experiments~\citep{Ballabriga:NIMA878:10,
   Delpierre:jinst9:C05059, Ballabriga:NIMA1045:167489, Procz:radmea127:106104,
   Darwish:ApplSpecRev55:243, Cheng:RSI93:013003, Unwin:commphys6:309}. In addition to single-event
time and location determination, its energy deposited in the sensor is also an important aspect.
Particularly, in AMO physics, the precise measurement of the charged-particle time-of-flight is
imperative for ionic fragments identification and the 3D momentum vector determination -- both
relevant for chemical dynamics studies~\citep{Ashfold:PCCP8:26, Chichinin:IRPC28:607}. Furthermore,
in the context of Bragg peak recognition for proton therapy, energy measurement makes it possible to
identify the energy deposition profile of protons, a key factor in dose
delivery~\citep{Paganetti:ProtonBeamTherapy:2017, Procz:radmea127:106104,
   Darwish:ApplSpecRev55:243}. In the realm of high-energy physics, energy information can inform
about the number of particles striking the detector simultaneously, contributing to particle
identification and event reconstruction~\citep{Aaij:JINST9:P09007, Akiba:NIMA723:47,
   Heijne:radmea140:106436, Bergmann:EPJC79:165}. Additionally, for astrophysics applications, the
brightness of events such as gamma bursts or while monitoring space weather, is invaluable to
determine the nature of the phenomena under investigation~\citep{Filgas:AstrN340:674}.

However, the question remains whether one can still accurately determine events when two or more
occur within a short period of time. This is particularly pertinent for AMO, high-energy physics and
radiation therapy applications, where high-intensity, high-frequency events are
commonplace~\citep{Heijne:radmea140:106436, Paganetti:ProtonBeamTherapy:2017,
   Bromberger:JPB55:144001}. Here, the detector's ability to accurately measure and differentiate
between rapid, successive events becomes crucial.

The Medipix collaboration has developed the Timepix line of hybrid-pixel
detectors~\citep{Poikela:JInst9:C05013, Llopart:NIMA581:485, Llopart:JINST17:C01044}. These
detectors, such as the current Timepix3 and Timepix4 models, offer high spatial and temporal
resolution, a low mass, low-power usage, high radiation tolerance as well as the possibility to be
used with visible photons~\citep{Llopart:JINST17:C01044, Heijhoff:JINST16:P08009,
   Garcia-Sciveres:RPP81:066101, FisherLevine:JInst11:C03016}. In this regard, the versions for
visible light have the great advantage that large-area detectors can be imaged onto the camera chip
using optical methods. Additionally, with the help of image intensifiers, a single photon
sensitivity can be achieved, important for instance for neutrino
experiments~\citep{Roberts:JInst14:P06001}. In general, all these detectors can measure single
particles at low noise and with a high dynamic range. They can time stamp these events by
time-of-arrival (ToA) at a resolution from 1.6~ns down to only hundreds of picoseconds. In addition,
the amplifier is designed such that the duration of the pulse produced by a hit is proportional to
the energy deposited in the sensor and thus, by measuring the duration of the pulse (time over
threshold, ToT) the deposited energy can be approximated~\citep{Poikela:JInst9:C05013,
   Llopart:JINST17:C01044, Pitters:JInst14:P05022}.

\begin{figure}
   \includegraphics[width=\linewidth]{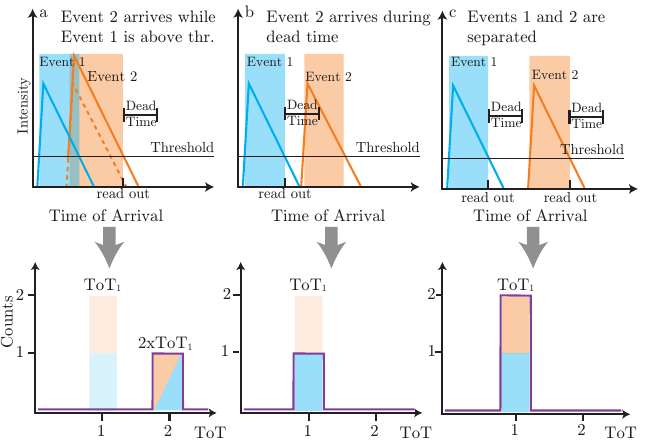}%
   \caption{Sketch of event processing in Timepix3. The top row presents an illustration of two
      events impinging on a detector pixel at different times. Their resulting intensity (ToT)
      histograms are drawn in the 2nd row, indicated by the purple line. The darker color represents
      the actual measured events, whereas the lighter colour indicates lost histogram events.}
   \label{fig:deadtime}
\end{figure}

The Timepix3 chip handles multiple consecutive events, illustrated in the upper row in
\autoref{fig:deadtime}, in three distinct ways depending on the timing of the incoming
events~\citep{Poikela:JInst9:C05013}, depicted in the lower row of \autoref{fig:deadtime}.
In \autoref[b]{fig:deadtime}, the second event arrives during the readout time of the first event
and, therefore, is discarded and only the first event is recorded. \autoref[a]{fig:deadtime} shows
the case when the second event arrives within the ToT of the first event or within 25~ns after the
signal from the first event went below the threshold. This results in a single detector event for
which, due to analogue pile up, the measured ToT is approximately the sum of the ToTs of the
individual events, as illustrated in \autoref[a]{fig:deadtime}. The third scenario is that the
second event arrives after the readout time of the first event is completed; both are recorded as
separate events. The expected ToT distribution, assuming equal intensities for both events, is
illustrated in the lower row and should consist of two hits with similar ToT values provided the two
events have the same amplitude.

Here, we report measurements on the influence of preceding pulses on the intensity recorded, which
demonstrate an influence for pulse-to-pulse separations up to $\ordsim10$~µs. Following a brief
description of our experimental setup, we present our experimental results from a systematic study
of the Timepix3 response to a second event in a single pixel within a few microseconds. We propose
an explanation for the observed effect, yielding a predictive model for Timepix3 that could be used
to correct the observed effect in post-analysis.

\section{Methods}
\label{sec:method}
\subsection{Experimental setup}
\label{sec:experimental-setup}
\begin{figure*}
   \includegraphics[width=\textwidth]{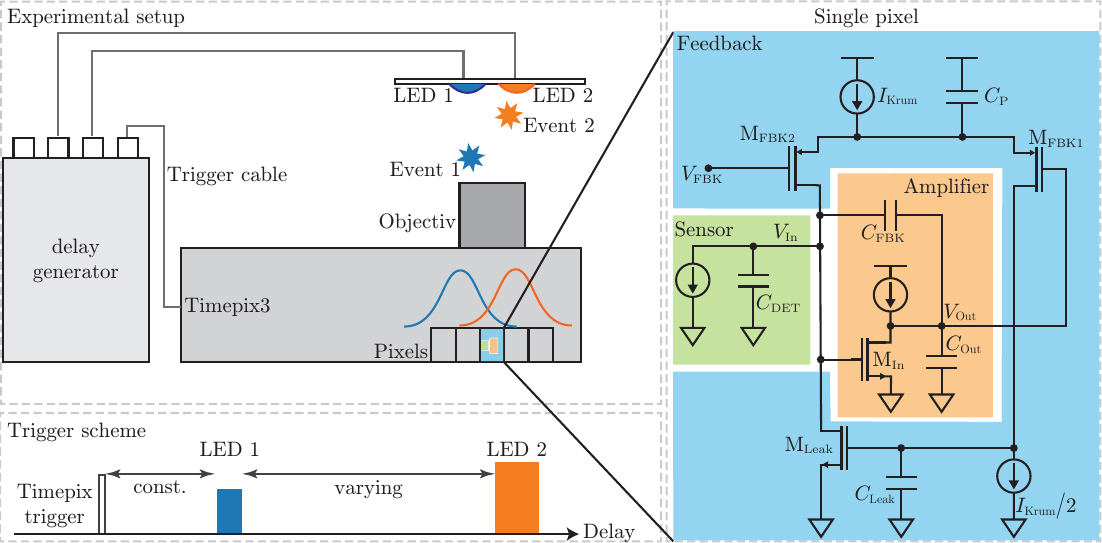}%
      \caption{Schematic of the experimental setup, the trigger scheme, and a single pixel circuit: The
      left upper part depicts the experimental setup with the Timepix3 camera facing two LEDs
      controlled for intensity and delays by a delay generator. The lower left part depicts the
      trigger scheme for the camera and LEDs, respectively. The height and width of the vertical
      boxes illustrate the control of the illumination for the respective LEDs by controlling the
      widths and amplitudes of the driver pulses. The right shows an illustration of a simplified
      schematic from a single pixel of the Timepix3-front-end amplifier, highlighting the equivalent
      small signal circuit model of a sensor (green), an amplifier circuit based on a cascaded
      common source transistor (orange), and the feedback reset path (blue). Currents, voltages, and capacitances are labelled by $I,V,C$ and
         transistors as M.}
   \label{fig:experimental_setup}
\end{figure*}
The experimental setup is illustrated in \autoref{fig:experimental_setup}. Two super-bright
light-emitting diodes (LEDs) are mounted next to each other facing a light-sensitive Timepix3 camera
built by Amsterdam Scientific Instruments based on homebuilt chips~\citep{FisherLevine:JInst11:C03016,
   Heijden:JINST12:C02040}. The camera objective is slightly out of focus to have the resulting
images of the two LEDs slightly overlap on the sensor. A delay generator (Stanford Research DG645)
was used to control the signal amplitude and relative delay of the LEDs. The delay generator was set
to internal trigger mode at a repetition rate of 100~Hz. The reference output channel of the delay
generator is used as a trigger-input signal for the Timepix3 camera.

To independently control the intensity and delay of the LEDs, each LED was connected to a separate
output channel of the delay generator. To operate the LEDs within a ToT range of 0.1--2~µs, the
channel level and pulse width were adjusted, as indicated in the lower part of
\autoref{fig:experimental_setup}, between $2\ldots5$~V and $15\ldots50$~ns, respectively. For the
smaller ToT the voltage was adjusted at a constant pulse duration of 15~ns until the desired ToT was
reached. All other ToTs were adjusted by picking the corresponding pulse duration at a channel level
of 5~V. All pulse durations applied were still short enough for the relevant timescales of the
experiment, \eg, some \us. For the delay scans, LED~1 was kept at a fixed delay with respect to the
reference trigger and the delay of LED~2 was swept.

The adjustable reference current \IKrum in the Timepix3 chip that controls how quickly the amplifier
output voltage in a pixel returns to its baseline value after charge is deposited was set to
$\IKrum=10$~steps; 1~step for the digital-to-analogue converter corresponds to 0.24~nA. The data was
recorded for 15~s, corresponding to 1500 event pairs, at each delay point. Additionally, all pixels
but one were electronically masked on the Timepix3 hardware. We carefully tested that the masking
does not influence the dynamics described below and is not influenced by effects that might be
related to synchronous illumination of a large fraction of the pixels in the matrix. The hardware was controlled using PymePix~\citep{AlRefaie:JInst14:P10003, pymepix:github} with
Tango~\citep{tango:website}.

To eliminate the possibility of the LEDs themselves being the source of the demonstrated
behavior, we also performed independent measurements of the
brightness of the two LEDs with a photodiode (Thorlabs, DET10). Here, the light of the LEDs was
focused with a 25~mm focal length lens onto the photodiode, the signal was amplified by
transimpedance amplifier (Femto, HVA-500M-20-B) and recorded by a digitizer (SPDevices ADQ14). The
amplitude of the second LED was constant within the error of the measurement and did not depend on
delay, as shown in \autoref{esi-fig:diode}.

\subsection{Circuit model}
To conceptionally understand the dynamics of a single Timepix3 pixel, it is required to look into
its electronics. A schematic of the simplified charge sensitive amplifier (CSA) circuit implemented
in every pixel used to amplify the signal delivered by the sensor is provided in the right part in
\autoref{fig:experimental_setup}. It illustrates (1) the equivalent small signal circuit model of a
sensor (green), (2) a simplified amplifier circuit based on a cascaded common source transistor
($\text{M}_\text{In}$) (orange) and (3) the feedback reset path (blue). The feedback reset path is
based on the Krummenacher architecture \citep{Krummenacher:nima305:527}, and it implements two
functions. The first function is to discharge the feedback capacitor ($C_\text{FBK}$) with a
constant current after the charge delivered by the sensor has been integrated. Since the discharge
rate is constant, it results in a ToT that is almost linearly proportional to the deposited charge.
This function is performed by the transistors $\text{M}_\text{FBK1}$ and $\text{M}_\text{FBK2}$
that, together with the tail-current source \IKrum, form a differential-pair structure. The second
function is to compensate for the leakage current produced by the sensor, which can vary over time.
The circuit can react to changes in leakage current at frequencies up to $\ordsim100$~kHz, though
usually the natural drifts occur more slowly than this. The leakage-current-compensation network is,
therefore, a low-pass frequency circuit in a feedback path whose purpose is to filter the very low
frequency components of the signal delivered by the sensor, \ie, the leakage current. The single
pixel circuit diagram includes $C_\text{FBK}$ and $C_\text{Leak}$ as real capacitance components.
$C_\text{Leak}$ is usually chosen on the order of hundreds of femto farads. The remaining capacitors
model equivalent or parasitic capacitances intrinsic to the electronics.

\subsection{Detector simulation}
\label{sec:method:simulation}
\begin{figure}[b]
   \includegraphics[width=\linewidth]{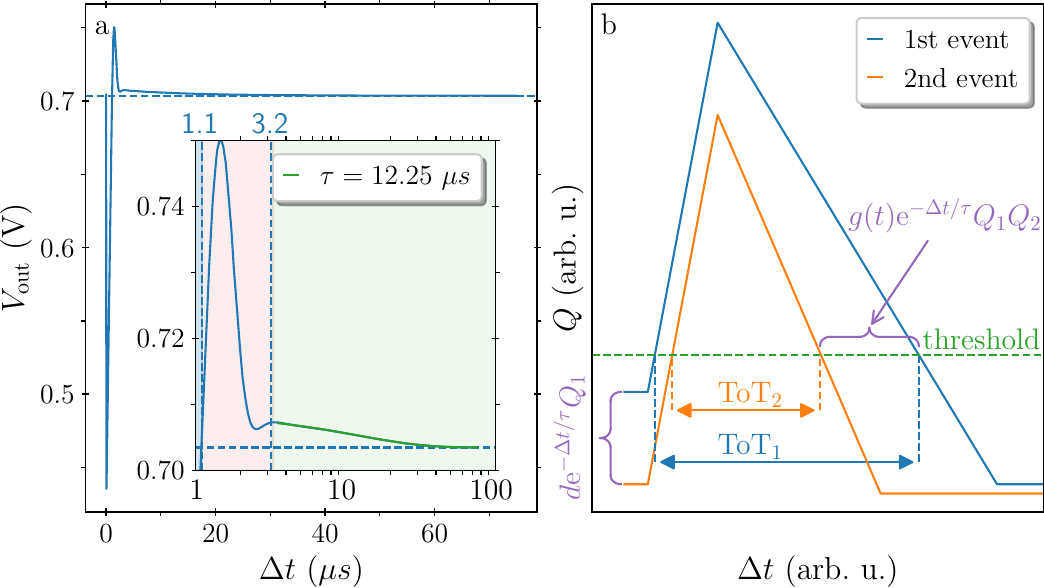}%
   \caption{(a) Response of the circuit to a 5000~electrons input impulse; note that the signal
      induced by the sensor has positive polarity. The inset depicts a zoom-in with a logarithmic
      time axis. (b) Illustration of the contributions for the presented loss model.}
   \label{fig:amp_resp}
\end{figure}
A simulation of the pixel response was made using Cadence Virtuoso~\citep{CadenceVirtuoso} based on
the circuit shown in \autoref{fig:experimental_setup}. \autoref[a]{fig:amp_resp} shows the output
voltage $V_\mathrm{Out}$ of the amplifier following an input pulse of 5000~electrons. The sensor
exhibits a positive polarity, which means that a decrease in $V_\mathrm{Out}$ is actually indicative
of an increase in input voltage $V_\mathrm{In}$. When the current pulse arrives, there is a rapid
initial voltage decrease (increase) within the first 50~ns for $V_\mathrm{Out}$ ($V_\mathrm{In}$).
Following this, there is a linear increase (decrease) in $V_\mathrm{Out}$ ($V_\mathrm{In}$), due to
the linear discharge of the capacitor $C_\text{FBK}$ through the feedback circuit. The duration of
the linear $V_\text{Out}$ increase is proportional to the magnitude of the input electron pulse.
This voltage is subsequently recorded by the discriminator to determine the ToT. The discharge
process spans approximately 1.4~µs, designated as region 1 (blue area), exhibits damped oscillations
in region 2 (red), and concludes slightly above $V_\text{Out}$ around 2.7~µs, marking the onset of
an exponential decay toward the initial voltage in region 3 (green); $V_\text{Out}$ converges with a
time constant of 12.3~µs.

\section{Results and Discussion}
\label{sec:results}
\subsection{Experimental results}
\label{sec:experimental-results}
\autoref[a]{fig:LED_SingleTrace} shows the measured \ToT{1} (blue dots) from LED~1 and \ToT{2}
(orange dots) from LED~2 as a function of the delay $\Delta t$ between the two LEDs. The brightness
was adjusted such that \ToTZ{1}$=0.710$~\us and \ToTZ{2}$=0.703$~\us. Furthermore, we set
$\IKrum=10$~steps.
\begin{figure}
   \includegraphics[width=\linewidth]{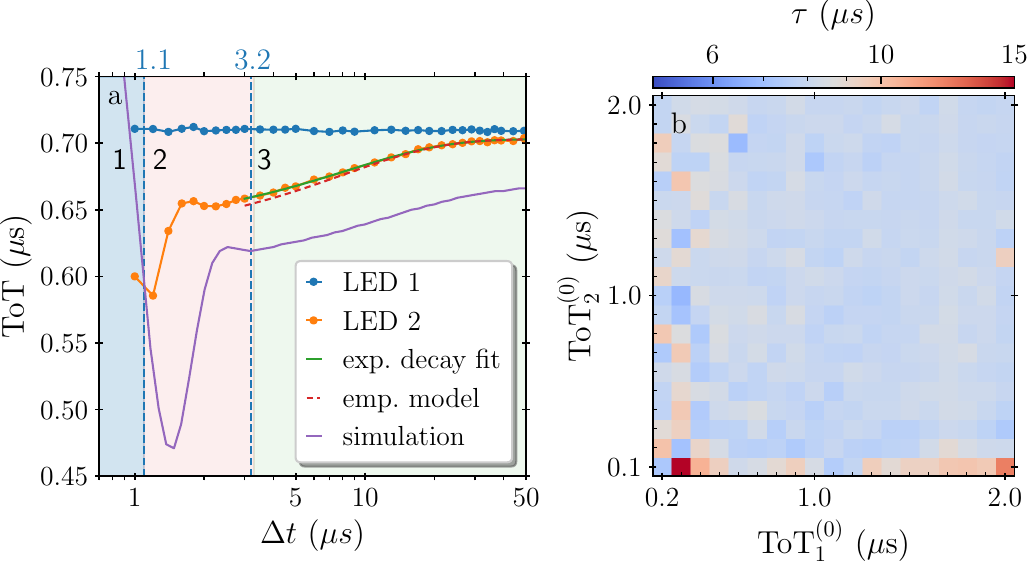}%
   \caption{(a) Example measurement of $\text{ToT}_\text{1}$ and $\text{ToT}_\text{2}$ as a function of the relative delay.
      Moreover, we show the fit of the exponential decay model to $\text{ToT}_\text{2}$ according to
      \eqref{eq:exp_model} (green solid line), the results from an electronic-circuit simulation
      (solid purple line), and the results from the empirical model (red dashed line). (b) Recovery
      times $\tau$ as a function of $\text{ToT}_\text{1}^{(0)}$ and $\text{ToT}_\text{2}^{(0)}$ obtained from the fits using
      \eqref{eq:exp_model}; see text for details.}
   \label{fig:LED_SingleTrace}
\end{figure}
\ToT{1} is constant (\ToTZ{1}) as a function of the delay whereas \ToT{2} shows a much richer
dynamics, and three distinct areas can be identified. In the first $\ordsim1.5$~µs (region~1) a
sharp drop of \ToT{2} is observed. \ToT{2} experiences a significant recovery over the next roughly
1.2~µs (region~2) and fully recovers within the following $\ordsim10$~µs to \ToTZ{2} (region~3). The
exponential recovery in the third region can be described by the model
\begin{equation}
   \text{ToT}_2(\Delta t) = \ToTZ{2}\left(1-B\exp(-\Delta t/\tau)\right)
   \label{eq:exp_model}
\end{equation}
with the recovering time $\tau$ and a fitting constant $B$. The green line in
\autoref[a]{fig:LED_SingleTrace} shows a fit to the data points with $\ToTZ{2}=0.703$~µs, $B=0.09$
and $\tau=8.4$~µs.

\autoref[b]{fig:LED_SingleTrace} shows the recovery time $\tau$ extracted from \eqref{eq:exp_model}
for a range of \ToTZ{1} and \ToTZ{2} combinations between 0.1 and 2~\us. The recovery time is almost
constant for all intensity combinations, with a mean recovery time of 8.2~µs and a root-mean-square
(RMS) deviation of 0.8~µs.

The relative signal, $\ToT{2}/\ToTZ{2}$, at $\Delta t=3~\us$ is shown in
\autoref[a]{fig:LED_RelativeSignal} as a function of \ToTZ{1} and \ToTZ{2}.
\begin{figure}
   \includegraphics[width=\linewidth]{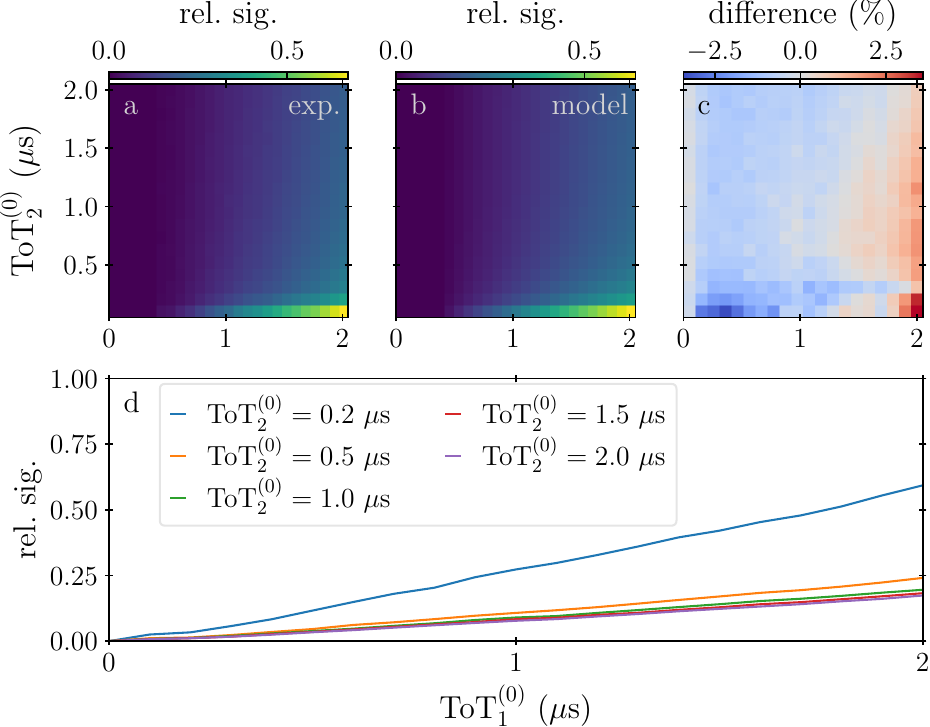}
   \caption{%
      (a) Experimental relative signal at a fixed delay of 3~µs given by
      $\text{ToT}_\text{2}(\Delta t=3~\us)/\text{ToT}_\text{2}^{(0)}$ from \eqref{eq:exp_model}. (b) Relative signal extracted
      from the empirical model. (c) Difference of the data in (a) and (b). (d) Lines presenting
      values from selected rows in panel (a) to visualize the quasi-linear dependence of the
      relative signal depending on $\text{ToT}_\text{1}^{(0)}$. See text for details.}
    \label{fig:LED_RelativeSignal}
\end{figure}
A linear increase as a function of \ToTZ{1} is observed for all \ToTZ{2}, which is also depicted in
\autoref[d]{fig:LED_RelativeSignal} for a selection of \ToTZ{2}-line outs. A larger slope is observed
for smaller \ToTZ{2} values. The linearity of the relative signal as a function of \ToTZ{1} for all
\ToTZ{1}--\ToTZ{2} combinations implies that the observed dynamics happens in the linear region of
sensor operation and is not a saturation effect caused by the first LED pulse.

\begin{figure}
   \includegraphics[width=\linewidth]{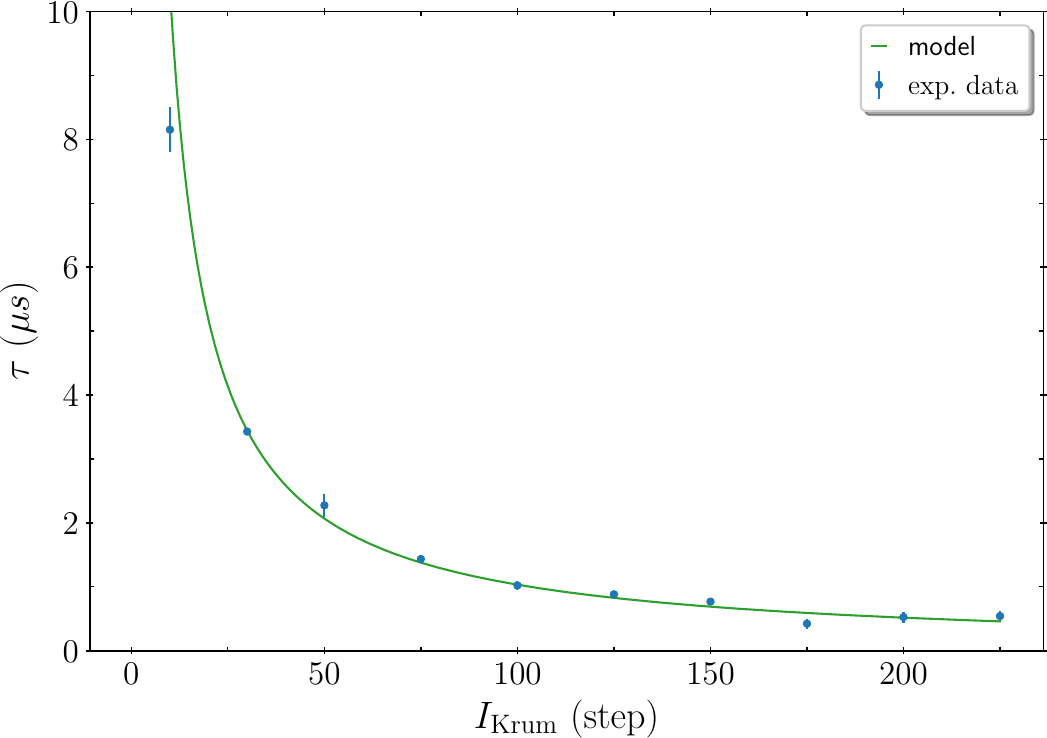}%
   \caption{Dependency of the recovery time $\tau$ on $I_\text{Krum}$. The line represents a weighted fit,
      $Q/I_\text{Krum}$ with $Q=103.4$, of the experimental data points with their standard
      error.}
   \label{fig:LED_IKrum}
\end{figure}

The relationship between the recovery time $\tau$ and \IKrum is illustrated in
\autoref{fig:LED_IKrum} as blue dots.
Empirically, it was found that the recovery time was inversely proportional to \IKrum. The fitting
model, described by $Q/\IKrum$ with $Q=103.4$, is represented by the orange line. The reciprocal
relationship between \IKrum and the recovery time implies that small changes in \IKrum at small
\IKrum will result in a significant change in the recovery time, whereas the asymptotic slope is
almost zero at higher \IKrum values, with $\tau=0.41$~µs at the maximum $\IKrum=250$~steps.

\subsection{Detector simulation}
\label{subsec:result:simulation}
As depicted in \autoref[a]{fig:LED_SingleTrace}, the variation of \ToT{2} with $\Delta t$ is closely
related to the amplifier's voltage response, and the non-equilibrium state of the amplifier upon
arrival of the second pulse will affect its measured ToT. The decrease in \ToT{2} is due to two
reasons, as depicted in \autoref[b]{fig:amp_resp}. Firstly, the non-equilibrium state in region 3
results in a larger current on the input side and, therefore, an effective larger \IKrum. This is
indicated by the steeper negative slope of the second event after the peak. Secondly, the shift in
the voltage results in a reduced pulse height above the threshold. This explains the
disproportionate reduction in signal observed when the second pulse is small, as seen in the
bottom-right of \autoref[a]{fig:LED_RelativeSignal}. In summary, both effects result in a decreased
\ToT{2} in region 3.

Careful inspection of the simulation results shows good qualitative agreement regarding the slow
recovery time extracted from the experiment. This can be explained by considering two components:
the capacitor $C_{\rm Leak}$ and the transistor $\rm M_{\rm Leak}$ in the leakage current
compensation circuit. The temporary increase in leakage current compensation due to an incoming
pulse is due to a charge increase on the capacitor $C_\text{Leak}$, which leads to a higher
$\rm M_\text{Leak}$ transistor gate voltage, increasing its current. This increased current in the
transistor in turn causes the capacitor to discharge, resulting in an exponential decay to the
baseline. The associated time constant is given by
\begin{equation}
    \tau_{\rm Leak} =\frac{C_{\rm Leak}}{\rm g_{\rm mM_{\rm Leak}}},
    \label{eq:tau_leak}
\end{equation}
where $\rm g_{\rm mM_{\rm Leak}}$ is the small-signal transconductance of the transistor
$\rm M_{\rm Leak}$. Given the values $C_{\rm Leak}\sim350$~fF and
$\rm g_{\rm m M_\text{Leak}}~24.5$~nS (for $\IKrum=1.5$~nA), this results in a time constant of
$\tau_\text{Leak}=14.3$~µs, which is close to the experimentally observed recovery time. A
qualitative sketch of the above described behaviour is provided in
\autoref{esi-fig:cir_charge-sketch}. As shown by \eqref{eq:tau_leak}, the time constant is inversely
proportional to the transconductance of the leakage current compensation transistor $\rm M_{Leak}$.
Experimentally, the time constant is inversely proportional to \IKrum, as shown in
\autoref{fig:LED_IKrum}, which implies that the transconductance increases linearly with \IKrum.
This is consistent with simulation; the transistor is being operated in weak inversion, and under
these conditions, the transconductance is proportional to the current.

So far, we have discussed region 3 of the amplifier- and \ToT{} response corresponding to timescales
longer than 3.2~µs. To account for the full temporal behaviour, we consider a circuit analysis using
the Laplace transform of the circuit's transfer function. Points where the Laplace transform tends
to infinity are called ``poles'', and correspond to characteristic responses of the circuit,
consisting generally of exponential responses, sinusoidal oscillations, or a combination of these
(\eg{} damped oscillations). The circuit contains five reactive elements (five capacitors) as
depicted in \autoref{fig:experimental_setup}. There are, however, only four independent initial
conditions for these elements because capacitors $C_\text{DET}$, $C_\text{FBK}$, and $C_\text{OUT}$
form a loop in which setting the initial conditions on two of them fixes the state of the third one.
This leads to a system with four poles, which are on the left half plane, \ie, they correspond to
either exponential decays or damped oscillations. One of these four poles corresponds to the long
exponential undershoot in region~3 already discussed above. A second one corresponds to the rising
amplifier pulse. Further detailed analysis of the circuit shows that, depending on the choice of the
dimensions of the transistors in the front-end and also on the parasitic capacitances that are
present in the circuit layout, the remaining two poles in the transfer function can be a pair of
complex conjugate poles, corresponding to a damped sinusoidal oscillation in the time domain. These
poles produce the pronounced damped oscillation of the time waveform signal in region 2. Therefore,
in region 2, a greater loss of the second pulse's \ToT{} is observed than expected from the
exponential decay alone, due to the sinusoidal undershoot suppressing the signal level. Finally, at
very short times below $\Delta t=1.1$~µs, when decreasing the delay time further, the second pulse
piles up in the tail of the first pulse, effectively ``decreasing'' the threshold, \ie,
``increasing'' the \ToT. Decreasing the delay time even further leads to a scenario when the two
pulses cannot be distinguished as separate events at the input of the discriminator, as explained in
\autoref[a--b]{fig:deadtime}.

The observed effect is intrinsic to this kind of amplifier design, but delicately depends on the
exact values of the used components and details of the circuit design used for the amplifier. We
also confirmed the behaviour with a different chip and multiple pixels, showing the overall same
characteristic with slightly different fitting constants. For example, as shown in
\autoref{esi-fig:tpx2_pixel1}, for another camera with an $\IKrum=10$, the recovery time was
$\tau\sim 4.5$~µs with a width of 0.4~µs. We attribute these seemingly different results to the fact
that the \IKrum value is not an absolute calibrated value across multiple sensors, but varies from
chip to chip.

\subsection{Loss model}
\label{sec:model}
To predict the expected reduction in \ToT{2} following an earlier illumination event with \ToT{1}
within delays observed in region 3, we developed an empirical mathematical model. Creating a model
that encompasses the more complex dynamics of regions 1 and 2 is outside the scope of this work; see
elsewhere for detailed numerical detector simulations~\cite{Ballabriga:JINST}. We assume that the
decrease of \ToT{2} was primarily influenced by two effects: (1) due to increased leakage
compensation a faster discharge of $C_\text{FBK}$ is obtained and (2) due to the too high discharge
of $C_\text{FBK}$ the voltage level at which the second pulse starts off is lower than in the
equilibrium level as illustrated in \autoref{fig:amp_resp}. With this, the relative loss
$r(\Delta t)$ can be modelled in first order as:
\begin{equation}
   1-\frac{\ToT{2}(\Delta t)}{\ToTZ{2}} \approx \ToTZ{1}\text{e}^{-
      \Delta t/\tau_\text{m}}\left(\frac{d}{\ToTZ{2}(\Delta t) - t_\text{ret}} + g\right).
    \label{eq:model}
\end{equation}
Here, \ToTZ{1} denotes the intensity of the first pulse in microseconds, $\Delta t$ represents the
time between the first and second hit in microseconds, $\tau_\text{m}$ is a fitting parameter
representing the recovery time in microseconds, $d$ is used to describe an offset between the
equilibrium leakage compensation and the one due to a higher discharge, $\ToT{2}(\Delta t)$ is the
measured amplitude of the 2nd event at the time $\Delta t$, $t_\text{ret}$ is an empirical parameter
to improve the fit and $g$ represents the factor responsible for the increased slope from a faster
discharge due to higher leakage compensation. The good agreement between model and experimental data
can also be observed in \autoref[a]{fig:LED_SingleTrace} indicated by the red dashed line. See the
corresponding \autoref{esi-sec:mmodel} in the supplementary
information for more details.

Applying the model to the data presented, we list the parameters as follows:
$\tau_\text{m}=8.03$~µs, $d=0.019$, $t_\mathrm{ret}=0.136$~µs, $g=0.112~\us^{-1}$. The model's
recovery time closely aligns with the experimental recovery time of $\tau=8.01$~µs, as seen in
\autoref[b]{fig:LED_SingleTrace}. Due to $t_\mathrm{ret}$ in the denominator, the model only works
for values greater than $\tau_\text{ret}$. Moreover, we note that due to variations in fabrication,
the value of these parameters will vary somewhat between different Timepix3 chips, \cf our
corresponding experimental findings above.

To further illustrate the strong agreement between our model and experimental data,
\autoref[c]{fig:LED_RelativeSignal} depicts the difference between the model's results outlined in
\autoref[b]{fig:LED_RelativeSignal} from \eqref{eq:model} and the actual experimental measurements
in \autoref[a]{fig:LED_RelativeSignal} at $\Delta t=3$~µs. Notably, the maximum relative difference of 3~\% between the model and the experimental data
is merely $\pm 3.5\,\%$, and this variation is most pronounced for the lowest \ToT{2}. The deviation
is attributed to the simplicity of the model. Comparing \autoref[a]{fig:LED_RelativeSignal} and
\autoref[b]{fig:LED_RelativeSignal}, this model allows estimating the real and correcting the
measured ToT. It consequently improves the time resolution and the intensity
RMS~\citep{Bromberger:JPB55:144001, Pitters:JInst14:P05022}.

To obtain this calibration for a given Timepix3 assembly, it is
beneficial to record more points for smaller $\text{ToT}_{2}$s, whereas for $\ToT{2}\gtrsim0.7$~µs
this is not as critical any more. For our data, we found it sufficient to measure \ToT{2} in the range
$0.1\ldots0.5$~µs in increments of $0.1$~µs and at $1,1.5,2$~µs to obtain a very good fit.

The real \ToTZ{2} can be obtained from the measured \ToT{2} and \ToT{1} by solving \eqref{eq:model}
for \ToT{2}. However, an easier way is to use the relative loss directly:
\begin{equation}
  \ToTZ{2} = \frac{\ToT{2}(\Delta t)}{1-r(\Delta t)}.
\end{equation}

\section{Conclusions}
\label{sec:conclusions}
Timepix3 based detectors not only enable precise nanosecond-level event timing, but also offer
valuable deposited-energy information, crucial for a wide range of applications and essential for
mitigating timewalk inaccuracies in time measurements. We found a systematic underestimation of
event intensity for events registered within $\ordsim10~\us$ after previous events in the same
pixel. For delays exceeding 2~µs, this effect diminished exponentially, with the time constant
contingent on $1/\IKrum$. Our comprehensive analysis, utilizing integrating circuit design and
simulations, is in very good agreement with the experimental observations.

Additionally, we introduce an empirical model to quantitatively assess the relative loss in the
second event when the time interval between the first and second events exceeds approximately 3~µs.
This model demonstrates exceptional alignment with experimental data and offers the potential for
real-time post-processing correction of ToT losses. The implementation of such correction measures
holds the promise of improving experimental data, thereby enhancing our proficiency in sub-pixel
positioning, timewalk correction, and related applications. To correct the data from other cameras
using the model, the above measurements are necessary in combination with the fits presented.

Overall, our findings not only advance our understanding of Timepix3's capabilities, but also
provide clear practical means to optimize the detectors' performance for a wide range of
applications that benefit from improved time resolution and precise event-intensity measurements. In
similar detectors like Medipix3 and Timepix4, the described effect is either not present or reduced
due to a different functionality or amplifier design, respectively. In future developments, it is
envisioned to eliminate this loss fully entirely.

\begin{acknowledgments}
   \section*{Acknowledgments}%
   We gratefully acknowledge Guido Meier for initial discussions of our findings and Amsterdam
   Scientific Instruments B.~V.\ for collaborative information sharing and discussions.

   \section*{Funding}
   We acknowledge financial support by Deutsches Elektronen-Synchrotron DESY, a member of the
   Helmholtz Association (HGF), also for the provision of experimental facilities and for the use of
   the Maxwell computational resources operated at DESY. We acknowledge support by the Matter
   innovation pool of the Helmholtz Association through the DataX project. The research was further
   supported by the Cluster of Excellence ``Advanced Imaging of Matter'' (AIM, EXC~2056,
   ID~390715994) of the Deutsche Forschungsgemeinschaft (DFG).
\end{acknowledgments}

\bibliography{string,cmi}

\onecolumngrid
\end{document}

\title{\textit{Supporting information:} \\
   Timepix3: single-pixel multi-hit energy-measurement behaviour}%
\author{Hubertus Bromberger}\cfeldesy%
\author{David Pennicard}\desy %
\author{Rafael Ballabriga}\cern \author{Sebastian Trippel}\stemail\cmiweb\cfeldesy\uhhcui%
\author{Jochen Küpper}\cfeldesy\uhhcui\uhhphys%
\date{\today}%
\maketitle

%\jknote{I get undefined references, see comment in source.}

%\tableofcontents
%\listoffigures
%\listoftables
%\onecolumngrid
%\bigskip\bigskip
%\twocolumngrid
%\clearpage

\subsection{Intensity measurements with a photodiode}
\begin{figure}[b]
   \includegraphics[width=\linewidth]{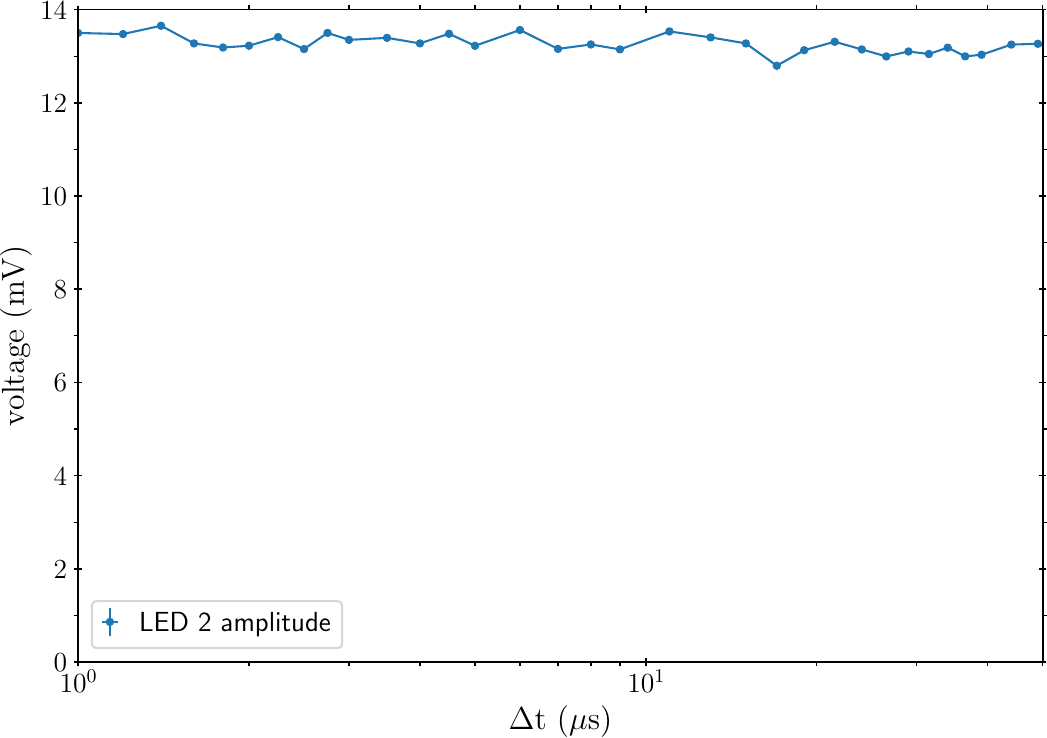}%
   \caption{Voltage amplitude of LED~2 measured with a photodiode as a function of the delay to
      LED~1; every point represents an average over 100 single measurements.}
   \label{fig:diode}
\end{figure}

To determine if the observed effect of a reduced \ToT{2} originates from the LEDs themselves, we
measured their intensity with a photodiode (Thorlabs, DET10) and an analog-to-digital converter
(SPDevices, ADQ14) in combination with a fast amplifier (Femto, HVA-500M-20-B). The measured voltage
which is proportional to the intensity is shown in \autoref{fig:diode} as a function of the relative
timing. A constant intensity is observed at all times. This shows that the observed exponential
recovery is not related to the LEDs and the light production themselves. The noise is attributed to
the low signal levels and the corresponding difficulties in determining the intensity.
% TODO: change markersize=3
% TODO: xlabel: t (µs)
% TODO: ylabel: voltage (mV)
% LED leer 2

\subsection{Recovery time comparison for different Timepix3 detectors}
\begin{figure}
      \includegraphics[width=\linewidth]{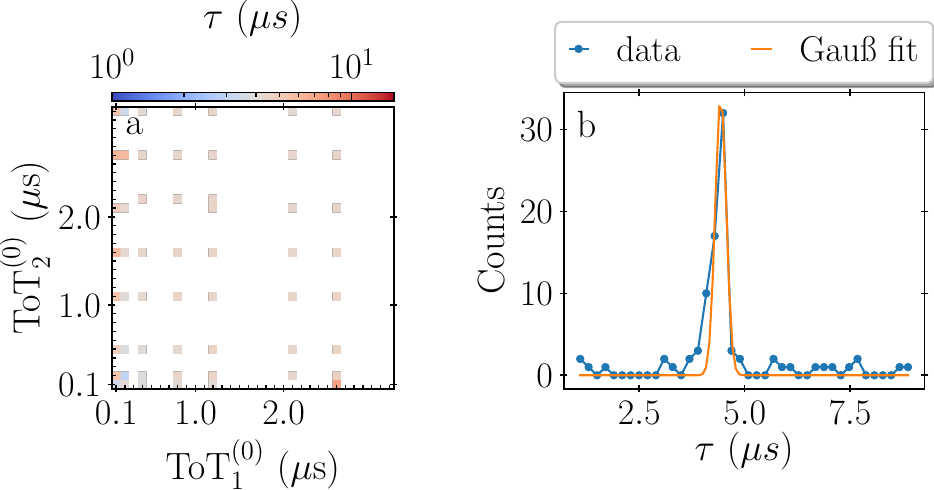}%
      \caption{Recovery time $\tau$ matrix from a different Timepix3 detector.}
      \label{fig:tpx2_pixel1}
\end{figure}

\autoref{fig:tpx2_pixel1} shows the recovery time for a more sparse matrix as provided in the main
manuscript for a different Timepix3 detector. \autoref[b]{fig:tpx2_pixel1} shows a histogram of the
corresponding recovery times and a Gaussian fit. The average recovery time was determined to be
$\tau=4.4$~µs with a $\rm FWHM=0.3$~µs. Although the \IKrum settings were the same as in the main
text, the recovery time is about half as long as for the other camera. This is attributed to the
fact, that the absolute values for \IKrum and the threshold are different across different
detectors.

\subsection{Temporal charge accumulation dynamics and regimes in sensor feedback circuit}
\begin{figure}
   \includegraphics[width=\linewidth]{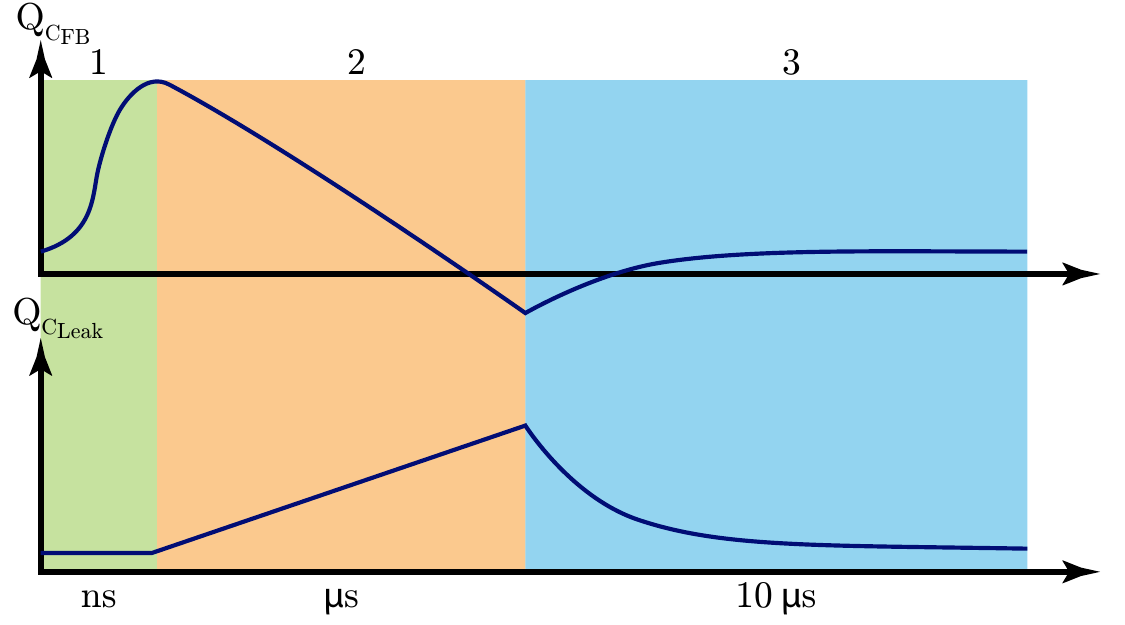}%
   \caption{Sketch of the temporal charge evolution for $C_\text{FBK}$ and $C_\text{Leak}$ after an
      incoming pulse.}
   \label{fig:cir_charge-sketch}
\end{figure}

The temporal charge accumulation after an incoming pulse from the sensor of $C_\text{FBK}$ and
$C_\text{Leak}$ is illustrated in \autoref{fig:cir_charge-sketch}. The three different regimes
encoded by colour correspond to the different time scales determined in the main text. During the
first phase, while the charge on the feedback capacitor $C_\text{FBK}$ is accumulated, the leakage
compensation capacitor is still not reacting. In the second phase, while $C_\text{FBK}$ is linearly
discharged, charge accumulates on $C_\text{Leak}$ in the same way. However, due to the lag in the
leakage compensation circuit, $C_\text{FBK}$ is discharged more than its equilibrium value. In phase
3, while the charge on $C_\text{Leak}$ returns to equilibrium, on the same timescale, $C_\text{FBK}$
also returns to its equilibrium charge state. See main text for a more detailed explanation.

\subsection{Mathematical model for loss}
\label{sec:mmodel}
In the following section, we provide a derivation of the developed model,
to estimate the loss in the second event during the exponential recovery phase. In first, order $\ToT{2}(\Delta t)$ is given by the following equation
\begin{align}
  \ToT{2}(\Delta t) &= \nonumber\\
  &\ToTZ{2} - d\text{e}^{-\Delta t/\tau_\text{m}} \ToTZ{1} - g\text{e}^{-\Delta t/\tau_\text{m}}\ToTZ{1}\ToTZ{2}.
\end{align}
The equation is based on the following arguments.
The initially accumulated charge from the second event and therefore \ToTZ{2} is reduced by two
components (a sketch is provided in \autoref[b]{main-fig:amp_resp} in the main document). Firstly, the
second term on the right side describes the reduction of \ToTZ{2} due to a lower voltage baseline in
a specific pixel. The lowering of the baseline should be proportional to the charge introduced by
the first pulse and therefore also in first order proportional to \ToTZ{1}. Additionally, the
baseline should recover to its equilibrium value exponentially with a time constant $\tau_\text{m}$.
The proportionality constant is given by $d$. Secondly, the third term on the right side describes
the reduction of \ToTZ{2} due to an increased slope for the charge flow. The increase of the slope
should be again proportional to \ToTZ{1} because it is caused by the first pulse. It should also
exponentially return to its equilibrium value with the same time constant $\tau_\text{m}$ as the
second term, since both effects are caused by the leakage current compensation as discussed in
\autoref{main-subsec:result:simulation} in the main document.  Furthermore, it should in first order linearly
dependent on the amount of charge introduced by the second pulse and therefore is again in first a linear function in \ToTZ{2}. This proportionality constant is given by $g$. Within this model, we can evaluate the relative loss,
$r(\Delta t)$, given by
\begin{align}
    r(\Delta t) &= 1-\frac{\ToT{2}(\Delta t)}{\ToTZ{2}} \nonumber\\
    &= \ToTZ{1}{\mathrm e}^{-\Delta t/\tau_\text{m}}\left(\frac{d}{\ToTZ{2}(\Delta t)} + g\right) \nonumber\\
    &= \ToTZ{1}{\mathrm e}^{-\Delta t/\tau_\text{m}}A. % TODO add ! on top of =
    \label{eq:model-1}
\end{align}
To obtain the fit constants $\tau_\text{m}, d$ and $g$ from the experiment and to empirically
potentially improve the expression, a series of fits were applied to the measured data. From
\autoref[d]{main-fig:LED_RelativeSignal} in the main document, the measured relative loss as a function of
\ToTZ{1} for various \ToTZ{2} is shown. A linear increase is observed for all \ToTZ{2} as expected
from \autoref{eq:model-1}. Their fitted slopes are depicted in \autoref[a]{fig:model_fit} as the
function of the relative timing $t$ between the two light pulses for various \ToTZ{2}. An
exponential decrease is observed for all \ToTZ{2} and the fitted time constants $\tau_\text{m}$ are
depicted in \autoref[b]{fig:model_fit} as a function of \ToTZ{2}. The time constant as a function of
\ToTZ{2} is constant ($\tau_\text{m}=8.03$~µs), which is in very good agreement with the data shown
in \autoref[b]{main-fig:LED_RelativeSignal} in the main document. \autoref[c]{fig:model_fit} shows the
remaining part $A$ corresponding to $d/\ToTZ{2}(\Delta t) + g$ in \autoref{eq:model-1} as a function of
\ToTZ{2}. The green curve is a fit to this functional dependency that does not fit the data
well, especially for small \ToTZ{2}. Therefore, an additional empirical retardation of \ToTZ{2} given by $t_{\mathrm{ret}}$ was
introduced, which leads to a good fit of the data as shown by the orange curve in
\autoref[c]{fig:model_fit}. The final model for the reduction in first order can therefore be written as
\begin{align}
    r(\Delta t) &= 1-\frac{\ToT{2}(\Delta t)}{\ToTZ{2}} \nonumber\\
    &= \ToTZ{1}{\mathrm e}^{-\Delta t/\tau_\text{m}}\left(\frac{d}{\ToTZ{2}(\Delta t)+t_{\mathrm{ret}}} + g\right).
    \label{eq:model-2}
\end{align}
Finally, \ToTZ{2} can directly be evaluated from the reduction using
\begin{equation}
  \ToTZ{2} = \frac{\ToT{2}(\Delta t)}{1-r(\Delta t)}.
\end{equation}
The complete procedure can again be found in the notebook, accompanied by the supplementary material.
\begin{figure}
   \includegraphics[width=\linewidth]{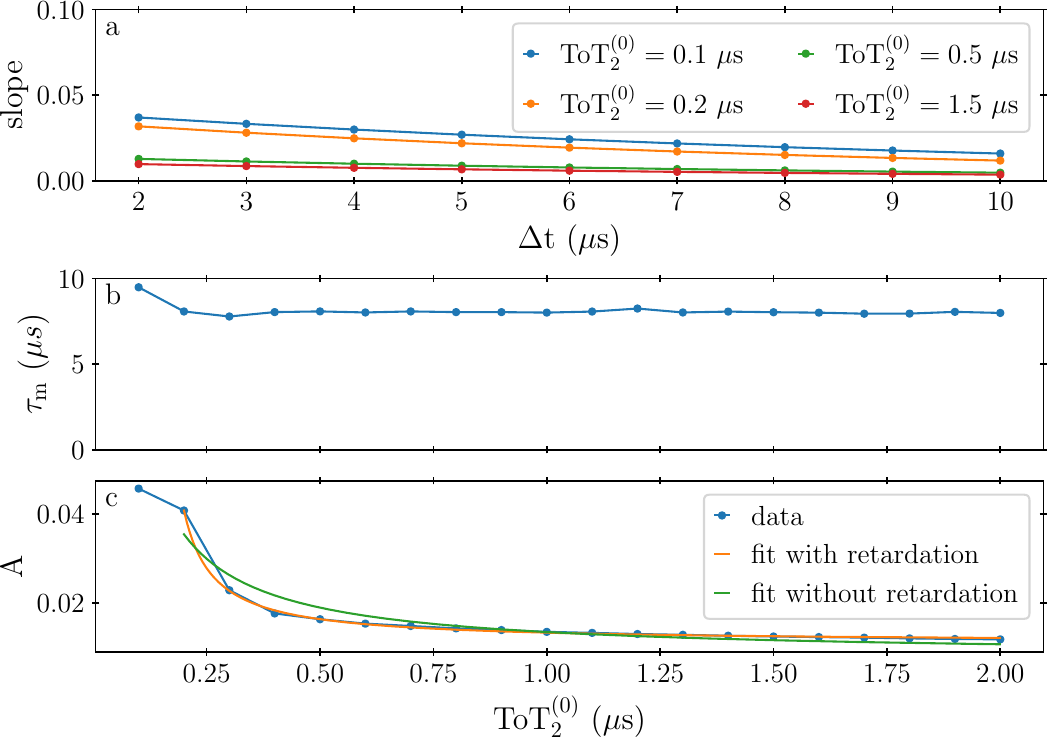}%
   \caption[Multiple steps for evaluating model fit parameters.] {%
      a) Slopes for different values of \ToTZ{2}. %
      b) Time constant $\tau_\text{m}$ as a function of \ToTZ{2}. %
      c) Amplitude of the exponential fit with and without retardation. See text for more detail.}
   \label{fig:model_fit}
\end{figure}
% TODO: ylabel: \tau_m
% TODO: ylabel: Amplitude -> A

\bibliography{string,cmi}

\newpage
\onecolumngrid
\listofnotes